# Giant spin Hall effects and topological surface states in ternary-layered MAX carbides $M_{n+1}AlC_n$ (M= Nb, Ta, n=1, 2, 3)

Yanhui Chen[1], Hong-Yan Lu[2], Wenjin Yang[1], Meifeng Liu[1], Bin Cui[3], Desheng Liu[3], Bing Huang[4], and Xi Zuo[1*]

[1]College of Physics and Electronic Science, Hubei Key Laboratory of Photoelectric Conversion Materials and Devices, Hubei Normal University, Huangshi 435002, China

[2]School of Physics and Physical Engineering, Qufu Normal University, Qufu 273165, China

[3]School of Physics, National Demonstration Center for Experimental Physics Education, Shandong University, Jinan 250100, China

[4]Beijing Computational Science Research Center, Beijing 100193, China

*Emails: zuoxi@hbnu.edu.cn

## Abstract

In this work, we report a systematic study of the electronic structures, band topology, and intrinsic spin Hall effect (SHE) of the layered MAX carbides $M_{n+1}AlC_n$ (M= Nb, Ta, n=1, 2, 3) and explore the correlation effects on the SHE. The results show that $M_3AlC_2$ and $M_4AlC_3$ (M= Nb, Ta) share similar Dirac-band-crossing features near the Fermi level ($E_F$) and form nodal lines in the absence of spin-orbit coupling (SOC). When the SOC is included, the Dirac band crossings are fully gapped, resulting in nontrivial $Z_2$ topological invariants (1;000) with a pair of surface states on the (001) plane. Remarkably, the multiple gapped Dirac points contribute to locally strong spin Berry curvatures, which lead to large spin Hall conductivities and a giant spin Hall angle up to ~ 60% for $Ta_3AlC_2$. Moreover, we also elucidate the impact of Hubbard U correction on SHC. Our findings indicate that $Ta_3AlC_2$ might represent an intriguing layered $Z_2$ topological metal with superior charge-to-spin conversion efficiency.

# I. INTRODUCTION

The spin Hall effect (SHE), a phenomenon driven by spin-orbit coupling (SOC) where a longitudinal electric current can induce a transverse spin current without an external magnetic field, has gained significant attention in recent years [1–4]. The SHE primarily arises from two mechanisms, namely, the extrinsic and intrinsic mechanisms [5]. The extrinsic mechanism arises from the defects and impurities [6–8], while the intrinsic mechanism originates from the topological nontrivial energy bands, which can be accurately evaluated by using the Kubo formula [9–11]. The spin Hall angle (SHA) of a SHE system is the ratio of the spin Hall conductivity (SHC) to the charge conductivity ($G_C$), which represents the charge-to-spin conversion efficiency at room temperature. Traditional SHE-based spintronic devices focus on heavy metals, such as fcc Pt [12] and $\beta$-W [13]. However, several issues, such as high experimental costs and relatively small SHA, limit their further development. Recently, with the rapid progress of topological states of matter, researchers have increasingly focused on topological materials such as topological insulators (TIs) and topological semimetals (TSMs), and explored the fundamental interplay between topological phases and SHE [14–17].

TIs are firstly reported in topological materials with strong SHE [18–20]. Theoretical study on several typical TIs, namely, $Sb_2Se_3$, $Sb_2Te_3$, $Bi_2Se_3$, $Bi_2Te_3$, demonstrates that the giant SHA (0.1-1.0) originates from the considerable intrinsic SHC (200-400 $((\hbar/e)(\Omega\,cm)^{-1})$) and small longitudinal conductivities [21]. However, their practical implementation is hindered by bulk carrier issues arising from the hybridization between surface and bulk states [22,23]. On the other hand, strong intrinsic SHE has also been found in TSMs such as the Dirac semimetals, Weyl semimetals, and nodal line semimetals [24–27]. Recently, Derunova et al investigate the intrinsic SHE of several A15 compounds and reveal giant SHC exceeding 1000 $(\hbar/e)(\Omega\,cm)^{-1}$ for $W_3Ta$, $Ta_3Sb$, and $Cr_3Ir$ [28]. Sattigeri et al extend the material candidate from Ta-based to Nb-based A15 compounds and demonstrate that these materials all belong to $Z_2$ topological metals with nontrivial Dirac surface states [29].

The giant SHC in the systems arises from the contribution of large spin Berry curvatures (SBC) around SOC-induced gapped nodal points and nodal lines (NLs). However, their pronounced metallic character gives rise to high $G_C$, which may constrain further enhancement of SHA. Therefore, a natural question arises whether we can identify ideal candidates in $Z_2$ topological metals possessing both large SHC and relatively small $G_C$, thus achieving giant SHA?

Recently, a group of layered hexagonal-structured ternary carbides, termed MAX compounds, has attracted much attention due to their remarkable combinations of the characteristics of metals and ceramics [30–32]. The general formula of MAX phases is $M_{n+1}AX_n$, where *M* represents an early transition metal, *A* stands for an early transition metal, *X* denotes C or N, and *n*=1, 2, 3. Interestingly, MAX phases can serve the precursors for the production of MXenes by selectively etching the "A" layer. This means that they both harbour a similar layered structure, and the various nontrivial topological insulating/semimetal phases reported in MXenes may also exist in MAX phases [33,34], which have been rarely explored. The above progress motivates us to study the electronic topology and SHE of MAX compounds.

In the paper, we perform a systematic study on the electronic structures, topological properties, and SHE of ternary-layered MAX carbides $M_{n+1}AlC_n$ (*M*= Nb, Ta, *n*=1, 2, 3). Firstly, the analysis of electronic band structures shows that all systems exhibit metallic characteristics with a strong $Z_2$ index (1;000) and nontrivial surface states. Secondly, we demonstrate that the SOC-induced gapped Dirac nodes induce strong SHC near the $E_F$. Accompanied with the inherently low $G_C$, the systems can achieve large SHA (1%-62%) at the $E_F$. Thirdly, we find that the magnitude of SHC components can be effectively tuned by introducing the Hubbard U. The rest of this article is organized as follows. In Sec. III A, we first give the description of crystal structures for ternary-layered MAX carbides $M_{n+1}AlC_n$ (*M*= Nb, Ta, *n*=1, 2, 3). In Sec. III B, we then report the band structures, topological invariants, and surface states for $M_{n+1}AlC_n$. In Sec. III C, we study the SHE and perform an analysis of band-decomposed and *k*-resolved spin Berry curvatures. In Sec. III D, we discuss the correlation effects on the SHE of Ta$_3$AlC$_2$. Finally, the conclusion drawn from this work is summarized in Sec. IV.

## II. THEORY AND COMPUTATIONAL DETAILS

First-principles calculations are carried out by using the density-functional theory (DFT) as implemented in the Quantum ESPRESSO package [35]. Projector-augmented wave (PAW) [36] and the generalized gradient approximation (GGA) with the Perdew-Burke-Ernzerhof (PBE) [37] functional are used to describe the potential of core electrons and the exchange-correlation interaction between the valence electrons, respectively. The kinetic-energy cutoff of the plane wave is set to be 100 Ry, and a 14 × 14 × 5 k-grid mesh of the Brillouin zone (BZ) is adopted in self-consistent calculations. The valence configurations of C, Al, Nb, and Ta atoms are $2s^22p^2$, $3s^23p^1$, $4s^25s^24p^64d^3$, and $5s^26s^25p^65d^34f^{14}$, respectively.

To explore the topological properties and SHE, the tight-binding Hamiltonians are constructed with the maximally localized Wannier functions for the outermost $p$ orbitals of C atoms, outermost $s$, $p$ orbitals of Al atoms, and outermost $d$ orbitals of Nb, Ta atoms generated by the first-principles calculations. The Wannier-fitted band structures are shown in Figs. S1 in the Supplemental Material (SM) [38].

Based on the tight-binding model constructed with WANNIER90, the NLs and topological surface states are calculated using the WannierTools software package [39,40].

The Kubo formula in the clean limit for SHC is given by

$$\sigma_{ij}^k = -\frac{e^2}{\hbar}\frac{1}{VN_k}\sum_n\sum_{\bm{k}} f_{n\bm{k}}\Omega_{n,ij}^k(\bm{k}), \tag{1}$$

$$\Omega_{n,ij}^k(\bm{k}) = \hbar\sum_{m\neq n}\frac{-2Im[\langle n\bm{k}|\hat{j}_i^k|m\bm{k}\rangle\langle m\bm{k}|\hat{v}_j|n\bm{k}\rangle]}{(E_{n\bm{k}}-E_{m\bm{k}})^2}, \tag{2}$$

where $\hat{v}_i = \frac{1}{\hbar}\frac{\partial H(\bm{k})}{\partial k_i}$, $\hat{j}_i^k = \frac{1}{2}\{\hat{s}_k, \hat{v}_i\}$ is the spin current operator, with the spin operator $\hat{s}_k = \frac{\hbar}{2}\hat{\sigma}_k$, and $i, j, k = x, y, z$. $|n\bm{k}\rangle$ is the eigenvector of the Hamiltonian $H$ corresponding to the eigenvalue $E_{n\bm{k}}$. $f_{n\bm{k}}$ is the Fermi-Dirac distribution for the $n$th band. $V$ is the primitive cell volume, and $N_k$ is the number of $k$ points sampled in the BZ. The unit of $\sigma_{ij}^k$ is $(\hbar/e)(\Omega\,\text{cm})^{-1}$. $\Omega_{n,ij}^k$ is referred to as the spin Berry curvature (SBC) in the unit of Å. For MAX carbides, a 100 × 100 × 100 Wannier interpolation $k$ mesh with 4 × 4 × 4 adaptive refinement $k$ mesh is used for the integral of the SHC.

The SHA is defined as the ratio of the SHC over the $G_C$, which characterizes the efficiency of converting the charge current to spin current. The SHA is evaluated according to

$$\theta_{\text{SH}} = \frac{2e}{\hbar}\frac{\sigma_{xy}^z}{\sigma_{xx}}, \tag{3}$$

where $\sigma_{xx}$ is the longitudinal $G_C$, $\sigma_{xy}^z$ is the transverse SHC. The longitudinal $\sigma_{xx}$ is calculated by using the Boltzmann transport equations within the constant relaxation time approximation as follows [41]:

$$[\sigma]_{ij}(\mu,T) = e^2 \int_{-\infty}^{+\infty}\left(-\frac{\partial f(\varepsilon,\mu,T)}{\partial(\varepsilon)}\right)\Sigma_{ij}(\varepsilon), \tag{4}$$

$$\Sigma_{ij}(\varepsilon) = \frac{1}{V}\sum_{n,\mathbf{k}} v_i(n,\mathbf{k})\,v_j(n,\mathbf{k})\tau(n,\mathbf{k})\delta(\varepsilon - E_{n,\mathbf{k}}), \tag{5}$$

where $\mu$ is the chemical potential, $f(\varepsilon,\mu,T)$ is the Fermi-Dirac distribution function $f(\varepsilon,\mu,T) = \frac{1}{e^{(\varepsilon-\mu)/k_B T}+1}$, $\Sigma_{ij}(\varepsilon)$ is the transport distribution function tensor, $E_{n,\mathbf{k}}$ is the energy of the $n$-th band at $\mathbf{k}$, $v_i(n,\mathbf{k})$ is the $i$-th component of the band velocity at $(n,\mathbf{k})$, $\delta$ is the Dirac's delta function, $V$ is the total volume of the system, and $\tau_{n,k}$ is the relaxation time depending on band and wave vector, which describes the collision term in the Boltzmann equation. In the calculation, we assume that the lifetime $\tau_{n,k}$ is independent of both n and k and choose the value $\tau = \tau_{n,k}$ by fitting the experimental electric conductivities at a given temperature.

## III. RESULTS

### A. Crystal structures

As shown in Fig. 1(a), the bulk $M_{n+1}AlC_n$ ($M$= Nb, Ta, $n$=1, 2, 3) crystallizes in a hexagonal crystal structure with the space group of P6$_3$/mmc (194), characterized by layered structures comprising alternating $M_{n+1}C_n$ slabs interleaved with planar Al atomic layers [42]. The carbon atoms occupy octahedral interstitial sites between adjacent Ta layers, forming edge-sharing $CTa_6$ octahedra. It is noted that the $M_3AlC_2$ and $M_4AlC_3$ studied here belong to the alpha phase, i.e., α-$M_3AlC_2$ and α-$M_4AlC_3$, in which the stacking sequence of Ta and Al atoms is AB**A**BAC**A**C and AB**A**BAC**B**CBC (the Al atoms are marked with bold letters), respectively [43,44]. As $n$ increases from 1 to 3, it can be seen that the height of c-axis increases linearly, indicating the addition of TaC layers. The optimized lattice parameters are given in Table S1 in the SM [38], which agrees well with previous experimental values.

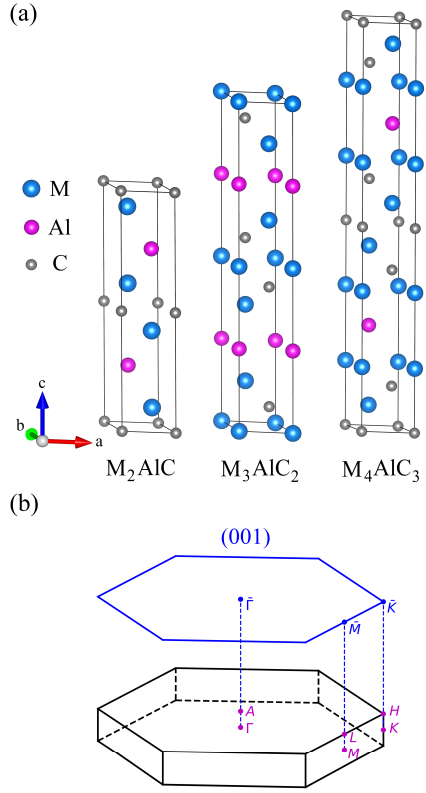

FIG. 1. (a) Crystal structure of $M_{n+1}AlC_n$ ($M$= Nb, Ta, $n$=1, 2, 3) compounds. (b) Brillouin zone (BZ) of a primitive cell of $M_{n+1}AlC_n$ and its projections towards the (001) surface.

## B. Electronic structures and topological properties of $M_{n+1}AlC_n$

Based on the optimized structures, we calculate the electronic structures of $Ta_{n+1}AlC_n$ ($n$=1, 2, 3) in Fig. 2, and those of $Nb_{n+1}AlC_n$ ($n$=1, 2, 3) are displayed in Fig. S2 in the SM [38], respectively. In Fig. 2(a)-(c), without considering the SOC, the red and upper blue bands near the $E_F$ touch with each other along Γ-M, K-Γ, and other routes, signifying its strong metallicity. The orbital character analysis shows that the 5$d$ orbitals of Ta dominate the bands near the $E_F$ (see Projected density of states in Fig. S3 in the SM [38]). Interestingly, $Ta_3AlC_2$ and $Ta_4AlC_3$ share very similar Dirac band crossings along K-Γ and A-L routes, indicating that they may harbor analogous topological properties. The inclusion of SOC significantly lifts the degenerate points and induces a continuous gap over the whole BZ (see yellow-shaded regions in Fig. 2(d)-(f)), and thus can be classified by the $Z_2$ topological invariant. Since the crystal structure possesses inversion symmetry, $Z_2$ topological invariants can be computed by evaluating the parity eigenvalues of all occupied bands at eight time-reversal invariant momenta (TRIMs), namely, Γ, A, three M, and three L points [45,46]. The calculated parity eigenvalues and $Z_2$ topological invariants ($v_0; v_1 v_2 v_3$) are listed in Table I. For all systems, we find that the $Z_2$ indices are (1;000), signifying their nontrivial topological character. The accuracy of $Z_2$ topological invariants is also verified by plotting the evolution of Wannier charge centers (WCCs) in the six time-reversal invariant momentum planes, which are shown in Fig. S4-S6 in the SM [38].

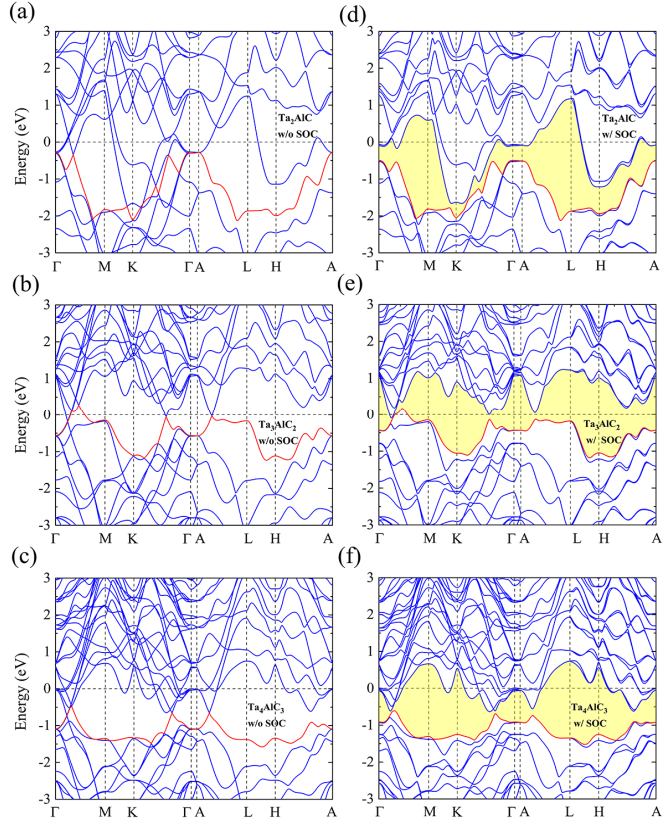

FIG. 2. Electronic structures without (a-c) and with (d-f) SOC of $Ta_{n+1}AlC_n$ ($n$=1, 2, 3). Light yellow regions in (d-f) represent the continuous band gaps between the red and blue bands.

To further examine the topological nontrivial nature, we calculate the surface states of $M_{n+1}AlC_n$ ($M$= Nb, Ta, $n$=1, 2, 3) on the (001) plane, as shown in Fig. 3. We first observe two nontrivial surface states (SSs) connecting the gapped Dirac points for all systems. Distinct from the localized SSs between $\bar{K}-\bar{M}-\bar{\Gamma}$ in $M_2AlC$ ($M$= Nb, Ta), the SSs are almost extended to the whole BZ in $M_3AlC_2$ and $M_4AlC_3$ ($M$= Nb, Ta). For $Nb_3AlC_2$ and $Ta_3AlC_2$, the SSs are well separated from the bulk states and located around the $E_F$, which should be easily detected by ARPES experiments. The nontrivial $Z_2$ indices and surface states demonstrate that $M_{n+1}AlC_n$ ($M$= Nb, Ta, $n$=1, 2, 3) are all $Z_2$ topological metals.

TABLE I. Parity products of all occupied bands at eight TRIMs and $Z_2$ topological indices in $M_{n+1}AlC_n$ ($M$= Nb, Ta, $n$=1, 2, 3).

| | MAX | Γ | A | 3M | 3L | $Z_2$ indices |
|---|---|---|---|---|---|---|
| $M_2AlC$ | $Nb_2AlC$ | + | − | − | − | (1;000) |
| | $Ta_2AlC$ | + | − | − | − | (1;000) |
| $M_3AlC_2$ | $Nb_3AlC_2$ | + | − | − | − | (1;000) |
| | $Ta_3AlC_2$ | + | − | − | − | (1;000) |
| $M_4AlC_3$ | $Nb_4AlC_3$ | + | + | − | + | (1;000) |
| | $Ta_4AlC_3$ | + | + | − | + | (1;000) |

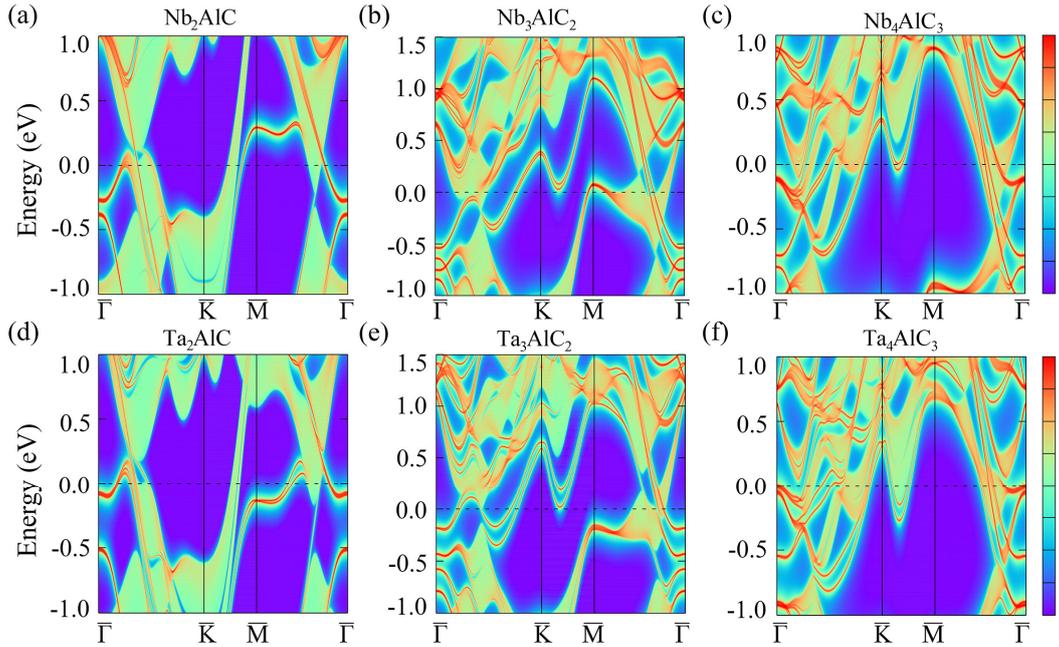

FIG. 3. Projected bulk bands onto the (001) surface for $M_{n+1}AlC_n$ ($M$= Nb, Ta, $n$=1, 2, 3).

## C. Spin Hall Effects

The multiple SOC-induced band anticrossing points aligned near the $E_F$ imply large spin Berry curvature (SBC) and strong SHE in $M_{n+1}AlC_n$. To explore SHE in $M_{n+1}AlC_n$, we first perform a symmetry analysis to confirm the allowed SHC components. As mentioned above, $M_{n+1}AlC_n$ has a hexagonal crystal structure with the space group of P6$_3$/mmc (194). The corresponding laue group is 6/mmm, leading to the constraints $\sigma_{xy}^z = -\sigma_{yx}^z$, $\sigma_{yz}^x = -\sigma_{xz}^y$, and $\sigma_{zx}^y = -\sigma_{zy}^x$, while other tensor elements are zero (see Table II) [47]. Therefore, there exist only three nonzero independent elements, namely, $\sigma_{xy}^z$, $\sigma_{yz}^x$, and $\sigma_{zx}^y$.

TABLE II. Symmetry-imposed tensor forms of the SHC tensors for $M_{n+1}AlC_n$ ($M$= Nb, Ta, $n$=1, 2, 3) with the space group of P6$_3$/mmc (194).

| | $\sigma^x$ | $\sigma^y$ | $\sigma^z$ |
|---|---|---|---|
| Space group P6$_3$/mmc | $\begin{pmatrix} 0 & 0 & 0 \\ 0 & 0 & \sigma_{yz}^x \\ 0 & \sigma_{zy}^x & 0 \end{pmatrix}$ | $\begin{pmatrix} 0 & 0 & \sigma_{xz}^y \\ 0 & 0 & 0 \\ \sigma_{zx}^y & 0 & 0 \end{pmatrix}$ | $\begin{pmatrix} 0 & \sigma_{xy}^z & 0 \\ \sigma_{yx}^z & 0 & 0 \\ 0 & 0 & 0 \end{pmatrix}$ |

The independent SHC components at the $E_F$ are shown in Table III. Firstly, all SHC components of Ta$_{n+1}$AlC$_n$ are larger than Nb$_{n+1}$AlC$_n$, which is in accordance with the enhanced SOC strength from Nb to Ta. Secondly, the magnitude of $\sigma_{xy}^z$ component is larger than other components for all systems, indicating the anisotropic properties in our system. This phenomenon can be explained by asymmetric structures in which the c-axis is much longer than the a- and b-axes. Thirdly, $\sigma_{xy}^z$ of Ta$_3$AlC$_2$ at the $E_F$ has the largest SHC value of -832 $(\hbar/e)(\Omega\,\text{cm})^{-1}$. This SHC value is larger than the intrinsic SHC in topological insulators Bi$_2$Te$_3 \sim 218$ $(\hbar/e)(\Omega\,\text{cm})^{-1}$ [21], and comparable to recent reported topological materials, such as ZrSiTe $\sim -755$ $(\hbar/e)(\Omega\,\text{cm})^{-1}$ [48] and MnBi$_2$Te$_4 \sim 845$ $(\hbar/e)(\Omega\,\text{cm})^{-1}$ at $E_F - 0.25$ eV [49]. In addition, the SHC components versus $E_F$ can be found in Fig. S7 in the SM [38].

TABLE III. Intrinsic SHC and SHA for independent tensor elements for bulk $M_{n+1}AlC_n$ ($M=$ Nb, Ta, $n=$1, 2, 3). The unit of SHC is $(\hbar/e)(\Omega\,cm)^{-1}$.

| MAX | | $\sigma_{xy}^z$ | $\sigma_{yz}^x$ | $\sigma_{zx}^y$ | $|\Theta_{xy}^z|$(%) | $|\Theta_{yz}^x|$(%) | $|\Theta_{zx}^y|$(%) |
|---|---|---|---|---|---|---|---|
| $M_2AlC$ | $Nb_2AlC$ | -191 | -44 | -103 | 1.12 | 0.26 | 3.26 |
| | $Ta_2AlC$ | -415 | -244 | -342 | 2.12 | 1.24 | 11.86 |
| $M_3AlC_2$ | $Nb_3AlC_2$ | -166 | -70 | -91 | 3.74 | 1.58 | 3.94 |
| | $Ta_3AlC_2$ | -832 | -317 | -298 | 62.32 | 23.8 | 27.8 |
| $M_4AlC_3$ | $Nb_4AlC_3$ | -185 | -94 | -82 | 2.78 | 1.42 | 3.52 |
| | $Ta_4AlC_3$ | -473 | -238 | -290 | 3.66 | 1.84 | 4.18 |

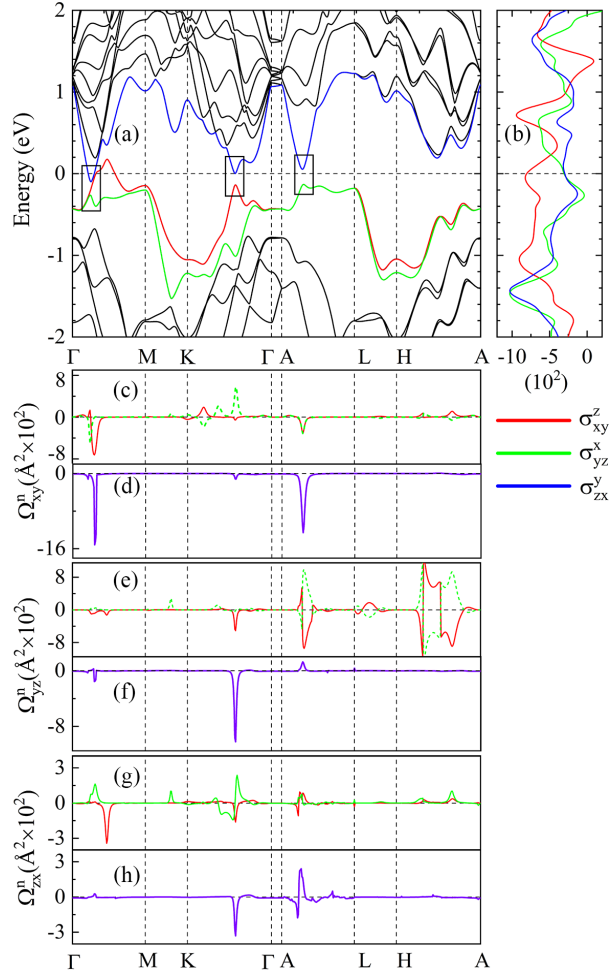

FIG. 4. $Ta_3AlC_2$ system. (a) Relativistic band structure; (b) spin Hall conductivities (SHC; $\sigma_{xy}^z$, $\sigma_{zx}^y$ and $\sigma_{yz}^x$) as a function of energy; (c), (e), (g) band-decomposed spin Berry curvatures (SBC, $\Omega^n$), as well as (d), (f), (h) total SBC along the high-symmetry lines in the Brillouin zone. In (a), (b), the $E_F$ is at zero energy, and the unit of SHC is $10^2\,(\hbar/e)(\Omega\,cm)^{-1}$. In (c)-(h), the unit of SBC is $Å^2$. Note that in (a), (c), (e), and (g), the same color curves correspond to the same bands.

On the other hand, we also show energy-dependent SHC ($\sigma_{xy}^z$) of Ta$_3$AlC$_2$ in Fig. 4(b) (see SHC components for other $M_{n+1}$AlC$_n$ compounds in Fig. S8-S13 in the SM [38]). It is observed that the magnitude of $\sigma_{xy}^z$ component reaches a local maximum right near the $E_F$, which is twice as large as that of $\sigma_{yz}^x$ and $\sigma_{zx}^y$ component. Moreover, $\sigma_{xy}^z$ component can still keep larger than -700 $(\hbar/e)(\Omega\,\text{cm})^{-1}$ for $E_F$ shifting down to $E_F - 0.4$ eV, and spikes to -900 $(\hbar/e)(\Omega\,\text{cm})^{-1}$ at $E_F - 1.0$ eV, while the $\sigma_{yz}^x$ and $\sigma_{zx}^y$ components reach -1000 $(\hbar/e)(\Omega\,\text{cm})^{-1}$ at $E_F - 1.4$ eV.

To further evaluate the charge-to-spin conversion efficiency, we calculate the SHA for our systems, which is defined a the ratio of the SHC value to the longitudinal $G_C$. The $G_C$ are calculated using the Boltzmann transport equations within the constant relaxation time approximation. According to the experimental conductivity values of $3.91 \times 10^4$ $(\Omega \cdot \text{cm})^{-1}$ [50] and $2.59 \times 10^4$ $(\Omega \cdot \text{cm})^{-1}$ [51] for Ta$_2$AlC and Ta$_4$AlC$_3$, the corresponding relaxation times are 6.65 and 10.20 fs, respectively. Due to similar geometry and electron structures, we assume that the relaxation times for Ta$_3$AlC$_2$ are equal to that of Ta$_2$AlC. The relation times for other systems and the calculated $G_C$ for $M_{n+1}$AlC$_n$ ($M$= Nb, Ta, $n$=1, 2, 3) can be found in Table SII and Table SIII in the SM [38], respectively. Here, we note that the calculated $G_C$ for Ta$_3$AlC$_2$ is $2.67 \times 10^3$ $(\Omega \cdot \text{cm})^{-1}$, which is an order of magnitude smaller than Ta$_2$AlC and Ta$_4$AlC$_3$, which is in accordance with the considerable band gaps at $E_F$ and smaller density of states reported in the literatures [50]. Therefore, the largest SHC values and smaller $G_C$ induce a giant SHA approaching 62.32% for Ta$_3$AlC$_2$, which is larger than fcc Pt (6.8%) [52], $\beta$−W (40%) [53], and topological insulators such as Bi$_2$Se$_3$ (26%) and Bi$_2$Te$_3$ (15%) [21].

To figure out the origin of large SHC in $M_{n+1}$AlC$_n$, we take Ta$_3$AlC$_2$ as an example to calculate the band-decomposed SBC at $E_F$ in Fig. 4(c)-4(h), and SBC analysis of other systems is shown in Fig. S8-S13 in the SM [38]. It is noted that the total SBC at $k$ in Fig. 4(d), 4(f), and 4(h) is the summation of SBC on all occupied bands at each $k$. According to previous studies, if a Dirac point opens a small hybridization gap with the inclusion of SOC at some $k$ point, then the SBC appears as a pair of peaks with opposite signs on upper and lower bands in the vicinity of this *k* point [28,48]. Therefore, when only one band is occupied. e.g., the $E_F$ falls within the gap, then only one peak of SBC would contribute to the total SHC.

As can be seen from Fig. 2(b) and Fig. 4(a), the inclusion of SOC lifts the degeneracies of all the band-crossing points between the red and blue bands, and partial band crossing points along the Γ-M routes between the green and red bands, as highlighted by black boxes. Correspondingly, these gapped nodal points induce prominent SBC peaks in Fig. 4(d), 4(f), and 4(h). We find that two giant SBC peaks along the Γ-M and A-L routes contribute to $\Omega_{xy}^z$ in Fig. 4(d), while only one large SBC peak along K-Γ contributes to the $\Omega_{yz}^x$ in Fig. 4(f). For $\Omega_{zx}^y$ in Fig. 4(h), apart from an obvious negative peak along K-Γ, there is another pair of peaks with opposite sign along A-L, resulting in a smaller SHC value compared to that of $\Omega_{xy}^z$. In addition, we also give the distribution of Dirac NLs and contour plots of $\Omega_{xy}^z(k)$ components in $k_z = 0$ in Fig. 5, which clearly shows that the SBC is mainly contributed by the gapped nodal points.

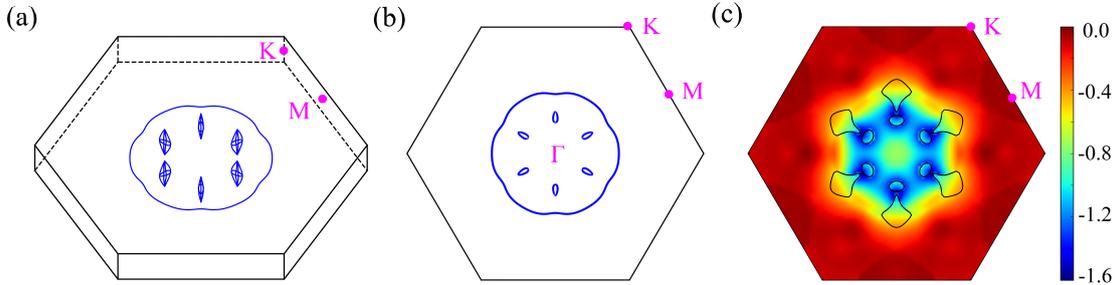

FIG. 5. (a) Distribution of NLs in the first BZ of Ta$_3$AlC$_2$ without SOC; (b) Top view of the NLs; (c) Contour plot of the spin Berry curvature component $\Omega_{xy}^z(k)$ in $k_z = 0$ plane.

### D. Correlation effects on SHE

Due to the 5d electrons inside the heavy metal elements Ta, the correlation effect is expected to have a strong influence on the electronic structures and SHE of MAX [54,55]. Taking Ta$_3$AlC$_2$ as an example, we introduce Hubbard U to evaluate the correlation effects on the electronic structures. Previous literature reports a Hubbard U parameter of 2.58 eV for Ta [56]. Therefore, we scan Hubbard U values from 0 to 3.0 eV in steps of 0.5 eV. We calculate the SHC as a function of Hubbard U in Fig. 6 for Ta$_3$AlC$_2$, and calculations of SHC versus Hubbard U for other systems (Ta$_2$AlC and Ta$_4$AlC$_3$) are given in Fig. S14 in the SM [38]. As the Hubbard U increases from 0 to 3 eV, we find that the absolute values of all three SHC components increase monotonically, with a total increase larger than 100 $(\hbar/e)(\Omega\,\text{cm})^{-1}$. For $\sigma_{xy}^z$, the maximum value is 950 $(\hbar/e)(\Omega\,\text{cm})^{-1}$ with a Hubbard U=3 eV.

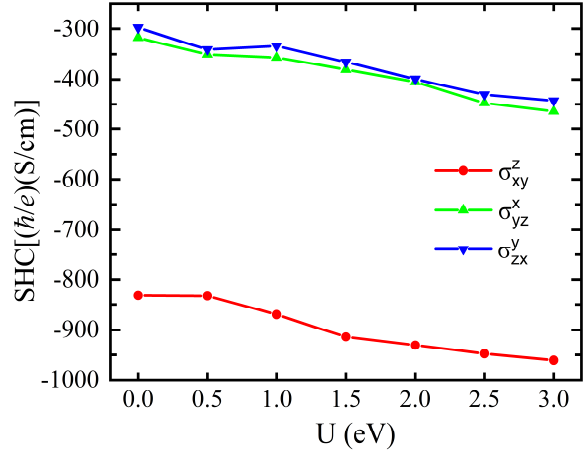

FIG. 6. SHC components as a function of U for Ta$_3$AlC$_2$.

To figure out the origin of the enhancement of SHC, we first give the electronic structures of Ta$_3$AlC$_2$ with SOC for U=1, 2, 3 eV in Fig. 7(a)-(c). As the U increases from 1 to 3 eV, we can see that Coulomb repulsion opens a larger band gap between the valence and conduction bands along the $k$-path. Then, we plot the contour plots of $\Omega_{xy}^z(k)$ and $\Omega_{yz}^x(k)$ components in $k_z = 0$ plane with different Hubbard U values. For the case of U=3eV in Fig. 7(f) and 7(i), the magnitude and distribution area of $\Omega_{xy}^z(k)$ and $\Omega_{yz}^x(k)$ is larger and more delocalized compared with other cases of U=1, 2 eV. According to the Kubo formula, the summation of SBC leads to higher SHC. In addition, it is noted that the positive correlation between the band gap and the SBC is quite common in recent papers on nodal line systems, such as $M$H$_2$ ($M$=Ti, Zr, Hf, Th) [57] and $M$X$_2$($M$=Th, Hf, Zr, X=Al, Ga, B) [58].

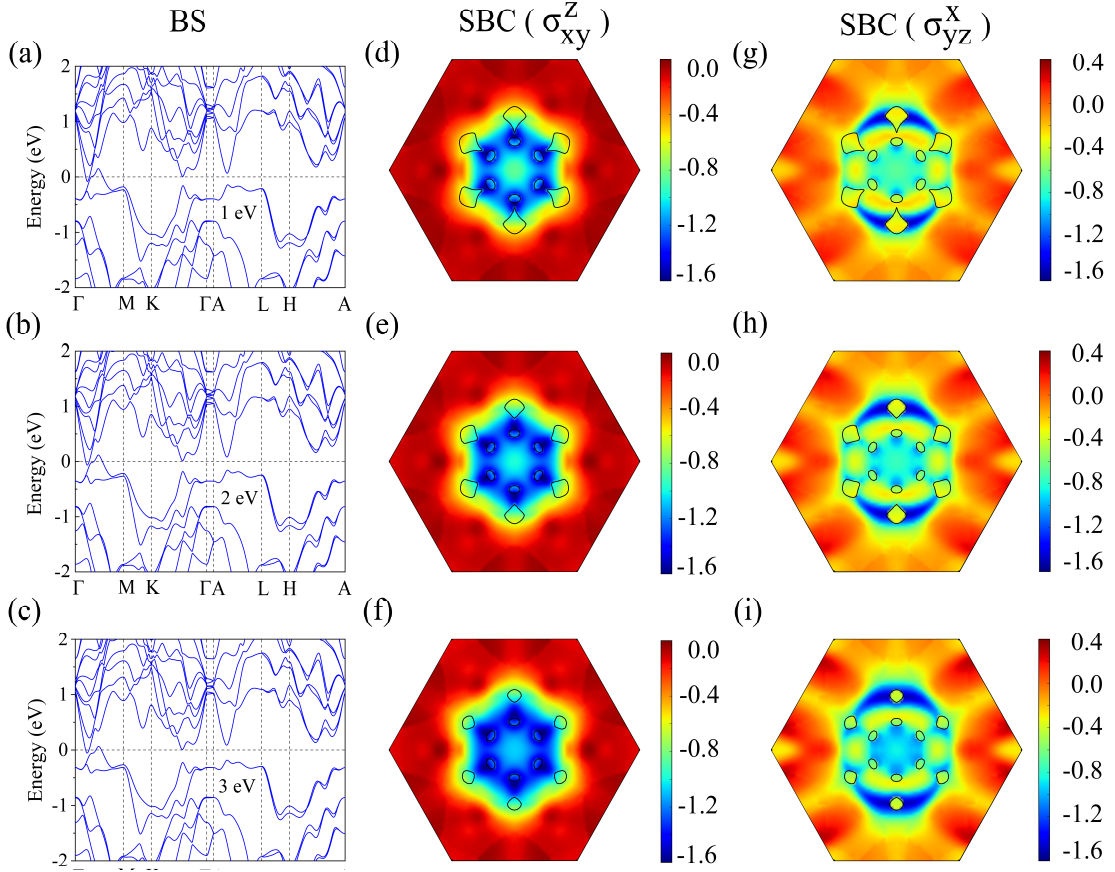

FIG. 7. Band structures (BS) of $Ta_3AlC_2$ with SOC under different Hubbard U values: (a) U = 1 eV, (b) U = 2 eV, and (c) U = 3 eV; (d)-(f) Corresponding $\sigma_{xy}^z$ compoment of SBC in first BZ; (g)-(i) Corresponding $\sigma_{yz}^x$ component of SBC in the first BZ. Note that the high-symmetry points are denoted as blue dots in (d)-(i).

## IV. DISCUSSION AND CONCLUSIONS

In conclusion, we have investigated the electronic structures, topological properties, and intrinsic SHE in ternary-layered MAX carbides $M_{n+1}AlC_n$ ($M$= Nb, Ta, $n$=1, 2, 3). All systems exhibit metallic characteristics with strong $Z_2$ indices (1;000) and nontrivial surface states. The SOC-induced Dirac anticrossing points aligned around $E_F$ lead to strong intrinsic SHE, with a large SHC of ~850 $(\hbar/e)(\Omega\ cm)^{-1}$ for $Ta_3AlC_2$ at the $E_F$. While the SHC of $M_{n+1}AlC_n$ is smaller than some topological metals, these materials exhibit a giant SHA (1%-62%) at the $E_F$. Furthermore, the inclusion of Hubbard U correction can arouse further enhancement of SHC in $Ta_3AlC_2$. Our results not only suggest the potential of layered MAX materials for advanced spintronic applications, but also elucidate the important role of Coulomb repulsion in regulating the intrinsic SHE in topological systems.

# ACKNOWLEDGMENTS

The authors thank Hong-Yan Lu for the helpful discussions. This work is supported by National Natural Science Foundation of China (Grant Nos. 12304139).